\documentclass[preprint,showpacs,preprintnumbers,amsmath,amssymb,aps,nofootinbib,,superscriptaddress]{revtex4}
\usepackage{bbm}
\usepackage{color}
\usepackage{mathrsfs}
\usepackage{amsmath,amssymb,bm,comment}
\usepackage{graphicx}
\usepackage{dcolumn}
\usepackage{bm}

\begin{document}

\title{Vorticity and magnetic field production in relativistic ideal fluids}

\author{Jian-hua Gao}
\email{gaojh@sdu.edu.cn}
\affiliation{Shandong Provincial Key Laboratory of Optical Astronomy and Solar-Terrestrial Environment, School of Space Science and Physics, Shandong University at Weihai, Weihai 264209, China}
 \affiliation{Key Laboratory of Quark and Lepton Physics (MOE), Central China Normal University, Wuhan 430079, China}
\author{Bin Qi}
\affiliation{Shandong Provincial Key Laboratory of Optical Astronomy and Solar-Terrestrial Environment, School of Space Science and Physics, Shandong University at Weihai, Weihai 264209, China}
\author{Shou-Yu Wang}
\affiliation{Shandong Provincial Key Laboratory of Optical Astronomy and Solar-Terrestrial Environment, School of Space Science and Physics, Shandong University at Weihai, Weihai 264209, China}


\begin{abstract}
In the framework of relativistic ideal hydrodynamics, we study the production mechanism  for vorticity and magnetic field in
relativistic ideal fluids. It is demonstrated  that in the uncharged fluids the thermal vorticity will always satisfy
the Kelvin's  theorem and the circulation must be conserved. However, in the  charged fluids,
the vorticity and magnetic field can be produced  by the interaction between  the entropy gradients and
the fluid velocity gradients. Especially, in the multiple charged fluids, the vorticity and magnetic field can   be produced
by the interaction between the  inhomogeneous charge density ratio  and the fluid velocity gradients even if  the entropy distribution is homogeneous,
which provides another   mechanism for the production of  vorticity and magnetic field in relativistic plasmas or in the  early universe.

\end{abstract}

\maketitle

\section{Introduction}
It is well known that the Universe is filled with vorticities and magnetic fields on all scales
\cite{Parker:1979,Carilli:2001hj,Ruzmaikin:1988,Zweibel:1997,Jones:1976tw}.
However the origin of these vorticities and magnetic fields is still one of the most challenging open problems in
theoretical physics \cite{Jones:1976tw,Widrow:2002ud,Kulsrud:2007an,Christopherson:2009bt}.
In the  nonrelativistic ideal fluid,
Kelvin's circulation theorem in hydrodynamics or
generalized version in magnetic hydrodynamics forbids the vorticity or magnetic fields to emerge from a
zero initial value when the fluid is  barotropic \cite{Landau:1987,Batchelor:1967,Roberts:1967}. In order to produce seed vorticity or magnetic field, we must resort to
the  baroclinic effects or go beyond the ideal fluids by including diffusive terms
\cite{Landau:1987,Batchelor:1967,Roberts:1967,DelSordo:2010mt,Dosopoulou:2011dg,Dosopoulou:2013zta}.
The presence of large-scale magnetic fields in the Universe,
especially the existence of an intergalactic magnetic field \cite{Kosowsky:1996yc,Harari:1996ac,Kosowsky:2004zh,
Kahniashvili:2008hx,Neronov:1900zz,Essey:2010nd,Pogosian:2011qv}, indicates the very possibility that the   magnetic fields
should  have been present in the early universe \cite{ Giovannini:2003yn,Giovannini:2006kg,Kandus:2010nw}, in which the temperature of the Universe is very high
and the velocities of the fluid and  the particle components are both relativistic. Hence we need relativistic hydrodynamics to deal with
these very hot fluid systems. Besides, relativistic hydrodynamics  is also a very important  theoretical
tool in high-energy heavy-ion physics. The ideal and dissipative hydrodynamics has succeeded
greatly in describing the collective flow from the data of RHIC and LHC.
The study of the vorticity and magnetic field production
in relativistic ideal fluids is very relevant to the important chiral effects, 
called the chiral-magnetic effects \cite{Kharzeev:2007jp,Fukushima:2008xe,Vilenkin:1979ui,Sadofyev:2010pr},
chiral-vorticity effects \cite{Vilenkin:1979ui,Kharzeev:2010gr}, and local polarization effects \cite{Landsteiner:2011cp,Gao:2012ix}, 
which can be expected in the noncentral heavy-ion collisions,  because all these effects  depend on the production of the  vorticity and magnetic field in the quark-gluon plasma.

In the relativistic ideal fluids,
there exists a similar covariant version of the Kelvin's circulation theorem \cite{Bekenstein:1978,Bekenstein:2000sf,Elsasser:2000xi}.
It turns out that  there are some subtleties  when we deal with the relativistic case, which have been pointed out in  Refs.\cite{Mahajan:2010, Mahajan:2011} that  vorticity and magnetic field can be produced in relativistic  purely ideal fluid due to space-time distortion caused by the special relativity.
However, all these relativistic  investigations up to date on the vorticity and magnetic field, as far as we know, are only limited to the systems
with single conserved charge and the  particle components in the fluids are also specified with finite mass.

It is well known that there exist the systems without any conserved charges theoretically or realistically, such as neutral $\phi^4$ field theory or photon gas.
In particular, the hydrodynamic simulation  used  in  relativistic heavy-ion collisions at RHIC or LHC are all based on the ones without charges and all the possible
charge imbalance is neglected \cite{Heinz:2013th,Gale:2013da,Teaney:2009qa,Romatschke:2007mq,Song:2007fn,Dusling:2007gi,Molnar:2008xj,Schenke:2010rr}.  
Hence it is very valuable to investigate both the  neutral fluids and charged fluids all together and
to see what could be missed only  from the neutral hydrodynamic equations and how the novel phenomenology could  appear in the single or multiple charged ones.

In this paper, we will extend these investigations to  more general
cases by direct manipulation of the relativistic hydrodynamic equations. We will not assume in advance that the particle components are massive or not
and the  systems we will consider can have  multiple conserving charges or    no conserving charge at all.
We find that the thermal vorticity will always satisfy the Kelvin's circulation theorem and be conserved in the uncharged fluids. However,
in the  charged fluids, especially in the multiple charged fluids, the vorticity and magnetic field can be produced not only by the interaction
between inhomogeneous entropy  and inhomogeneous fluid velocity magnitude but also by the interaction between inhomogeneous charge density ratio
and inhomogeneous fluid velocity magnitude. The latter provides another new  mechanism for the production of  vorticity and magnetic field in the
early universe or in the quark gluon plasma produced in heavy-ion collisions at RHIC or LHC.

\section{Vorticity in Relativistic ideal uncharged fluids }

In this section, we consider relativistic fluids without any conserving current,  in which  the hydrodynamical equations are just the energy-momentum conservation
\begin{eqnarray}
\label{div-Tmn}
\partial_\nu T^{\mu\nu}&=&0,
\end{eqnarray}
where $T^{\mu\nu}$ is the energy-momentum tensor. In the ideal hydrodynamics, $T^{\mu\nu}$ can be decomposed into
the following form,
\begin{eqnarray}
\label{Tmunu}
 T^{\mu\nu}&=&\left(\varepsilon+P\right) u^\mu u^\nu - P g^{\mu\nu},
\end{eqnarray}
where $\varepsilon$ is the energy density in the local frame, $  P$ is the pressure of the
fluid, the metric tensor $g^{\mu\nu}$ is chosen as $(1,-1,-1,-1)$, and  the fluid 4-velocity $u^\mu=\left(\gamma,\gamma{\bm v}\right)$ with the relativistic kinematic factor
$\gamma=1/\sqrt{1-{ v}^2}$ and   the normalization $u^2=1$.
Substituting Eq.(\ref{Tmunu}) into Eq.(\ref{div-Tmn}) and  contracting both sides with fluid velocity $u^\mu$, we can have
\begin{eqnarray}
\label{div-Tmn-t}
 u^\nu \partial_\nu\varepsilon+ (\varepsilon+{  P}) \partial_\nu u^\nu &=&0.
\end{eqnarray}
With  the general  equations from thermodynamics,
\begin{eqnarray}
\label{ds}
T ds &=& d\varepsilon,\ \\
\label{s}
T s &=& \varepsilon +P,
\end{eqnarray}
it is easy to verify that Eq. (\ref{div-Tmn-t}) is just the entropy current conservation,
\begin{eqnarray}
\label{entropy}
 \partial_\mu \left(s u^\mu\right) &=&0.
\end{eqnarray}
Using Eq. (\ref{s}), we can rewrite the energy-momentum tensor as
\begin{eqnarray}
\label{Tmunu-1}
 T^{\mu\nu}&=&T s u^\mu u^\nu - P g^{\mu\nu}
\end{eqnarray}
With the entropy conservation (\ref{entropy}), the energy-momentum conservation can be rewritten by
\begin{eqnarray}
\label{div-Tmn-e}
s u^\nu \partial_\nu \left(T u^\mu\right) -\partial^\mu {  P} &=&0.
\end{eqnarray}
Using the Gibbs relation $dp=sdT,$
we can have the following identity,
\begin{eqnarray}
\label{div-Tmn-e-1}
u^\nu \partial_\nu \left(T u^\mu\right) -\partial^\mu { T} &=&0.
\end{eqnarray}
It is  convenient to define the antisymmetric thermal vorticity tensor
\footnote{It should be clarified that there exists in  Refs. \cite{Becattini:2013vja,Becattini:2013fla}
another definition of the thermal vorticity from $u^\mu/T$  instead of $T u^\mu$ here. }
$\Xi^{\mu\nu}$ by
\begin{eqnarray}
\Xi^{\mu\nu}&=& \partial^\nu \left( T u^\mu\right) -  \partial^\mu \left( T u^\nu\right),
\end{eqnarray}
which is in complete analogy to the electromagnetic field tensor and represent both inertia and thermal forces.  With such a definition,
we can rewrite Eq.(\ref{div-Tmn-e-1}) as
\begin{eqnarray}
\label{T-EM-2}
\Xi^{\mu\nu}u_\nu=0
\end{eqnarray}
The circulation of the 4-vector temperature current $T u^\mu$ along the covariant loop $L(s)$ where
$s$ denotes the proper time is given by
\begin{eqnarray}
\label{T-EM-3}
\frac{d}{ds}\oint_{L(s)} T u^\mu dx_\mu = \oint_{L(s)} \Xi^{\mu\nu}u_\nu dx_\mu =0
\end{eqnarray}
which is just the relativistic Kelvin circulation theorem.
For a specific observer, the  vorticity  is always
defined in a fixed frame; hence, we need to consider the circulation of 3-vector temperature current $T \gamma \bm v$
along the synchronic loop $L(t)$.
We specify the components of thermal vorticity tensor $\Xi^{\mu\nu}$  in the  three-dimensional space as
\begin{equation}
\Xi^{\mu\nu}=
\left(  \begin{array}{cccc}
    0 & -{\cal E}^1 & -{\cal E}^2 & -{\cal E}^3 \\
    {\cal E}^1 & 0 & -{\cal B}^3 & {\cal B}^2 \\
    {\cal E}^2 & {\cal B}^3 & 0 & -{\cal B}^1 \\
    {\cal E}^3 & -{\cal B}^2 & {\cal B}^1 & 0 \\
  \end{array} \right),
\end{equation}
with the 3-vector definition,
\begin{eqnarray}
\label{calE}
 { \cal\bm E}&=&\left({\cal E}^1, {\cal E}^2 ,{\cal E}^3\right)
 = \left[\bm \nabla \left(T\gamma\right)-\partial_t \left(T\gamma \bm v\right)\right],\nonumber\\
 {\cal\bm B} &=& \left({\cal B}^1, {\cal B}^2 ,{\cal B}^3\right)=\bm \nabla \times \left(T\gamma \bm v\right)
\end{eqnarray}
With the above definition, we can  express the space components of Eq. (\ref{T-EM-2})  as
\begin{eqnarray}
\label{T-4}
{\cal\bm E} + \bm v \times {\cal \bm B} &=& 0
\end{eqnarray}
or
\begin{eqnarray}
\label{div-Tmn-s-s-2}
\partial_t \left(\frac{T {\bm v}}{\sqrt{1-{\bm v}^2}}\right) +\left(\bm \nabla \times \frac{T {\bm v}}{\sqrt{1-{\bm v}^2}}\right)
\times \bm v &=&- \bm \nabla { \frac{ T}{\sqrt{1-{\bm v}^2}}}
\end{eqnarray}

It should be noted that all through our paper the spatial hypersurfaces are always defined by the observer
in the lab frame instead of the comoving frame with velocity $u^\mu$.
Using the above identity, we can immediately obtain the conservation of the thermal current circulation in synchronic space,
\begin{eqnarray}
\label{VM-noc}
\frac{d}{dt}\oint_{L(t)} { T \gamma \bm{v}}\cdot d \bm x &=&
\oint_{L(t)} \left[\partial_t \left(\frac{T {\bm v}}{\sqrt{1-{\bm v}^2}}\right) +\left(\bm \nabla \times \frac{T {\bm v}}{\sqrt{1-{\bm v}^2}}\right)
\times \bm v \right]\cdot d\bm x\nonumber\\
&=&-\oint_{L(t)} \left[\bm \nabla { \frac{ T}{\sqrt{1-{\bm v}^2}}} \right]\cdot d\bm x=0
\end{eqnarray}
which implies that the thermal vorticity cannot emerge from a zero initial value.
It should be noted that, although the circulation of  the thermal current $ T \gamma \bm{v}$   is conserved, that of the kinetic current
$\gamma \bm{v}$ can be not conserved generally when the temperature is inhomogeneous.
By applying the Stokes theorem, the conservation of circulation can be transformed into the conservation
 of the  flux  ${\cal B}$ through the  surface which moves along with the fluid
\begin{eqnarray}
\frac{d}{dt}\int_{S(t)} { \cal{B}}\cdot d \bm S=\frac{d}{dt}\oint_{L(t)} { T \gamma \bm{v}}\cdot d \bm x =0
\end{eqnarray}
which means that the vorticity field flux is conserved or the vorticity lines are frozen in.

It should be noted that our result here cannot be naively regarded as a particular case of Ref. \cite{Mahajan:2010,Mahajan:2011}
 because all the derivations in \cite{Mahajan:2010,Mahajan:2011} are based on nonzero charge density. Once we set the charge density to vanish,
 we need another derivation from the beginning. This is actually what we are devoted to do in this section.

\section{Vorticity and Magnetic fields in Relativistic ideal magnetohydrodynamics with multiple currents}
In this section, we are devoted to discussing the relativistic fluids with multiple conserved currents. There are good reasons
to investigate the hydrodynamics with multiple currents. For example, in the hot and dense QCD matter produced in heavy-ion
collisions at high energy, one should be able to introduce electric charge, baryon number, and strangeness into the system.
Therefore multicharge hydrodynamics is important in developing hydrodynamic models in heavy-ion collisions. As we already mentioned in the
Introduction,  the hydrodynamic simulations  used  in  heavy-ion collisions at RHIC or LHC are all based on the ones without
charges \cite{Heinz:2013th,Gale:2013da,Teaney:2009qa,Romatschke:2007mq,Song:2007fn,Dusling:2007gi,Molnar:2008xj,Schenke:2010rr}, 
and so it is very valuable to  go beyond the  neutral fluids and investigate
how the novel phenomenology could  appear in the single or multiple charged ones.
 Besides, in the early universe, the quantum or thermal fluctuations between different charges  such as leptonic charge, electric charge, baryonic charge,
  and so on cannot coincide with each other; hence, it will be very important
 to investigate if such incoincidence could contribute to the production of vorticity or magnetic fields in the early universe.

Now let us assign one of the currents to  the
electric current $J^\mu$  from the local gauge symmetry, which can interact with the fluids by the magnetohydrodynamic equations. 
The other $m$ currents $J_i^\mu (i=1,2,...,m)$ are from the global symmetry, such as the baryonic current, leptonic current, and so on.
Then the magnetohydrodynamic equations for such a system are given by
\begin{eqnarray}
\label{MH-T}
\partial_\nu T^{\mu\nu}&=& F^{\mu\nu}J_\nu,\\
\label{MH-J}
\partial_\mu J^\mu&=&0,\\
\label{MH-JA}
\partial_\mu J_i^\mu&=&0,\ \  (i=1,2,...,m)
\end{eqnarray}
where $F^{\mu\nu}$ is the electromagnetic  stress  tensor and
can be written in terms of the electromagnetic 4-potential $ A^\mu $
as $ F^{\mu\nu}=\partial^\mu A^\nu -\partial^\nu A^\mu$.

The constitutive equations for the ideal magnetohydrodynamics read
\begin{eqnarray}
\label{T-EM}
T^{\mu\nu}&=&(\varepsilon+P) u^\mu u^\nu - P g^{\mu\nu}\\
\label{J-EM}
J^\mu&=& n u^\mu\\
\label{JA-EM}
J_i^\mu&=& n_i u^\mu,
\end{eqnarray}
where $n$ is the electric charge density and $n_i$ is charge density corresponding to
 other global symmetry.
The energy-momentum conservation (\ref{MH-T}) and current conservation (\ref{MH-J}) can yield
\begin{eqnarray}
\label{T-EM-1}
 n u^\nu \partial_\nu \left(\frac{\varepsilon+P}{n}u^\mu\right) -\partial^\mu P=n F^{\mu\nu}u_\nu
\end{eqnarray}
We can define the  generalized  thermal vorticity tensor $\Xi^{\mu\nu}$ by
\begin{eqnarray}
\Xi^{\mu\nu}&=& F^{\mu\nu}+\partial^\nu \left( f u^\mu\right) -  \partial^\mu \left( f u^\nu\right)
\end{eqnarray}
with $f=(\varepsilon+P)/{n}$.
We can rewrite Eq.(\ref{T-EM-1}) as
\begin{eqnarray}
\label{T-EM-2-n}
n \partial^\mu f -\partial^\mu P=n \Xi^{\mu\nu}u_\nu
\end{eqnarray}
Using the thermal equation
\begin{eqnarray}
T ds &=& d\varepsilon -\mu d n -\sum_{i}\mu_i d n_i, \\
T s&=& \varepsilon +P - \mu n -\sum_i \mu_i n_i
\end{eqnarray}
we can have the Gibbs relation corresponding to the multiple charge components,
\begin{eqnarray}
\label{Gibbs}
Td\left(\frac{s}{n}\right)
&=&d\left(\frac{\varepsilon}{n}\right)+Pd\left(\frac{1}{n}\right)-\sum_i\mu_i d\left(\frac{n_i}{n}\right)
\end{eqnarray}
where $\mu$ and $\mu_i$ denote the chemical potentials with respect to different conserving charges.
Now we can rewrite Eq.(\ref{T-EM-2-n}) as
\begin{eqnarray}
\label{T-EM-3}
\Xi^{\mu\nu}u_\nu &=& T\partial^\mu \left(\frac{s}{n}\right) +\sum_i \mu_i \partial^\mu \left(\frac{n_i}{n}\right)
\end{eqnarray}
It follows that the circulation of the 4-vector  current $f u^\mu +A^\mu $ along the covariant loop $L(s)$ is given by
\begin{eqnarray}
\label{T-EM-3}
& &\frac{d}{ds}\oint_{L(s)} \left(f u^\mu +A^\mu\right) dx_\mu = \oint_{L(s)} \Xi^{\mu\nu}u_\nu dx_\mu \nonumber\\
&=&\oint_{L(s)}\left[T\partial^\mu \left(\frac{s}{n}\right) +\sum_i \mu_i \partial^\mu \left(\frac{n_i}{n}\right)\right] dx_\mu
\end{eqnarray}
where $f u^\mu +A^\mu$ can be regarded as the canonical momentum or minimal coupling prescription (for details see \cite{Mahajan:2003}).
It is obvious that the circulation of this 4-vector  current is conserved  when $T$ and $\mu_i$ are constant.  Just like we
did in the last section, we need to consider the vorticity circulation of the synchronic loop $L(t)$. Let us define the 3-vector,
\begin{eqnarray}
\label{calE}
 { \cal\bm E}&=&{\bm E}+ \left[\bm \nabla \left(f\gamma\right)-\partial_t \left(f\gamma \bm v\right)\right],\nonumber\\
 {\cal\bm B} &=& {\bm B} + \bm \nabla \times \left(f\gamma \bm v\right)
\end{eqnarray}
Then the space components of Eq.(\ref{T-EM-3}) can be written as
\begin{eqnarray}
\label{T-EM-4}
\gamma\left({\cal\bm E} + \bm v \times {\cal \bm B}\right) &=& T\bm\nabla \left(\frac{s}{n}\right)
+\sum_i\mu_i {\bm \nabla}\left(\frac{n_i}{n}\right).
\end{eqnarray}
It follows that
\begin{eqnarray}
\label{VM}
& &\frac{d}{dt}\oint_{L} { \left(f\gamma \bm{v}+\bm A\right) }\cdot d \bm x=\frac{d}{dt}\int_S { \cal{B}}\cdot d \bm S\nonumber\\
&=&-\int_S \left[ \bm\nabla \left(\frac{T}{\gamma}\right)\times
\bm \nabla \left(\frac{s}{n}\right)\right]\cdot d\bm S
-\sum_i \int_S \left[ \bm\nabla \left(\frac{\mu_i}{\gamma}\right)\times
\bm \nabla \left(\frac{n_i}{n}\right)\right]\cdot d\bm S,
\end{eqnarray}
where $\bm{A}$ is the spatial part of $A^{\mu}$.
The second line  of the above equation is the source term which can lead to the vorticity or magnetic fields
from the zero initial value.
If we set $\mu_i=n_i=0$, we will recover the results obtained in Ref. \cite{Mahajan:2010},
\begin{eqnarray}
\label{VM-one}
& &\frac{d}{dt}\oint_{L} { \left(f\gamma \bm{v}+\bm A\right) }\cdot d \bm x=
-\int_S \left[ \bm\nabla \left(\frac{T}{\gamma}\right)\times
\bm \nabla \left(\frac{s}{n}\right)\right]\cdot d\bm S
\end{eqnarray}
As pointed out in Ref. \cite{Mahajan:2010}, the source term can be decomposed into the usual baroclinic term,
\begin{eqnarray}
S_b\equiv-\int_S \left[\frac{1}{\gamma} \bm\nabla T\times
\bm \nabla \left(\frac{s}{n}\right)\right]\cdot d\bm S,
\end{eqnarray}
and the pure relativistic term,
\begin{eqnarray}
S_r\equiv-\int_S \left[T\bm\nabla \left( \frac{1}{\gamma}\right) \times
\bm \nabla \left(\frac{s}{n}\right)\right]\cdot d\bm S,
\end{eqnarray}
which is absent in the nonrelativistic limit. When the velocity  and  entropy gradients are comparable, the baroclinic term
can be neglected  in the highly relativistic region due to the estimate \cite{Mahajan:2010}
\begin{eqnarray}
\frac{|S_{r}| }{|S_{b}|}\approx \frac{v^2}{1-v^2}
\end{eqnarray}

Now when  the multiple currents are involved, we notice that an extra new term,
\begin{eqnarray}
S_n\equiv-\sum_i \int_S \left[ \bm\nabla \left(\frac{\mu_i}{\gamma}\right)\times
\bm \nabla \left(\frac{n_i}{n}\right)\right]\cdot d\bm S,
\end{eqnarray}
arises. This is the principal result of this paper. It is very interesting that this term will generate the vorticity or magnetic field
even when the entropy is homogeneous where the first term in the second line  of Eq.(\ref{VM}) will vanish.  This new term can be broken into
two terms too; one is
\begin{eqnarray}
\label{Snmu}
S_{n\mu}\equiv-\sum_i \int_S \left[\frac{1}{\gamma} \bm\nabla \mu_i\times
\bm \nabla \left(\frac{n_i}{n}\right)\right]\cdot d\bm S
\end{eqnarray}
and the other is
\begin{eqnarray}
\label{Snr}
S_{nr}\equiv-\sum_i \int_S \left[\mu_i \bm\nabla \left( \frac{1}{\gamma}\right) \times
\bm \nabla \left(\frac{n_i}{n}\right)\right]\cdot d\bm S
\end{eqnarray}
Following the similar argument for $S_r$ and $S_b$ above, when the velocity  and  chemical potential gradients are comparable,
$S_{n\mu}$ term can be neglected  in the highly relativistic region due to
\begin{eqnarray}
\frac{|S_{nr}| }{|S_{n\mu}|}\approx \frac{v^2}{1-v^2}.
\end{eqnarray}
 Therefore, the dominant contribution will be
from the $S_{nr}$ term.

Compared with the results obtained in Refs.\cite{Mahajan:2010,Mahajan:2011}, in which only the single conserved current is included,
we have considered multiple currents in our work and  obtained new contributions when the ratio of the different charge densities
is inhomogeneous. This provides another possible  mechanism for the production of  vorticity and magnetic field in relativistic plasmas
or in the  early universe.

\section{Discussion and conclusion}

First, we  emphasize  that our result Eq.(\ref{VM-noc}) for  the ideal fluid without conserving current cannot be derived from the
result Eq.(\ref{VM}) with conserving currents  by naively taking the limit of $n \rightarrow 0$ and $n_i \rightarrow 0$ because there exists
the $1/n$ term. Take the neutral $\phi^4$ field or photon gas as examples. In these systems there is no
conserved charge at all, and we cannot introduce the charge density from the beginning.
That is why we must consider the ideal fluid without conserving currents separately. Although the result Eq.(\ref{VM-one}) with single
current can be found in the literature everywhere, we failed to find the result Eq.(\ref{VM-noc}) in the literature. Hence we
have  given the derivation of the result Eq.(\ref{VM-noc}) in our paper in Sec.II. 
The result reveals that  the thermal vorticity  always satisfies the Kelvin's circulation theorem and  cannot emerge from a zero initial value.

With respect to the result for the multiple currents in Eq.(\ref{VM}), the contribution from  the terms of $S_n$ or $S_{n\mu}$ and
 $S_{nr}$ is new. These terms, especially the $S_{nr}$ term, are very relevant to the production of vorticity or magnetic fields in the early universe,
where the particles can carry different charges, such as leptonic charge, electric charge, baryonic charge, and so on. Besides, it is very relevant
to the quark gluon plasma produced in heavy-ion collisions. If there is any inhomogeneous local distribution for some different charges, the vorticity or magnetic fields will be induced through the mechanism in Eq.(\ref{Snr}).  Then the chiral-magnetic effects,
chiral-vorticity effects,  and local polarization effects
 \cite{Kharzeev:2007jp,Fukushima:2008xe,Vilenkin:1979ui,Sadofyev:2010pr,Kharzeev:2010gr,Landsteiner:2011cp,Gao:2012ix}
will follow  in the noncentral heavy-ion collisions.

The above result is very relevant to the recent investigation on the chiral vortical effect  at RHIC.
The baryon-number separation observed by STAR \cite{Xu:2014jsa} can be explained by the chiral vortical effect; however,
such interpretation to a great extent depends  on the existence of large vorticity. In the Bjorken scaling scenario \cite{Bjorken:1982qr},
which is widely used as the initial condition for the hydrodynamic equations in  the relativistic heavy-ion collisions, the initial vorticity
must be zero. The results in our present paper show that the vorticity will always be zero when we insist on  using the ideal hydrodynamic
equations without any conserved currents and the chiral vortical effect does not appear at all. If we still want to be  in the regime of
ideal hydrodynamics, in order to estimate the possible chiral vortical effect, we must resort to the hydrodynamic equations with single or
multiple conserved currents.

\begin{acknowledgments}
The authors thank the referee for providing  valuable comments and help in improving the contents of this paper.
J.H.G. thanks Zhang-Bu Xu for  fruitful discussion.
J.H.G.  was supported in part by  the Major State Basic Research Development Program in China (Grant No. 2014CB845406), 
the National Natural Science Foundation of China under  Grant No.~11105137 and  CCNU-QLPL Innovation Fund (QLPL2014P01).
B.Q. was supported in part by the National Natural Science Foundation of China under  Grant No.~11005069 and
S.Y.W. was supported in part by the National Natural Science Foundation of China under  Grant No.~11175108.

\end{acknowledgments}

\end{document}